\documentclass[conference]{IEEEtran}
\IEEEoverridecommandlockouts
\usepackage{cite}
\usepackage{amsmath,amssymb,amsfonts}
\usepackage{algorithmic}
\usepackage{graphicx}
\usepackage{multirow}
\usepackage{threeparttable}
\newcommand{\TblFont}{\footnotesize}
\newcommand{\TblScale}{0.96}
\newcommand{\TblColSep}{3pt}
\newcommand{\TblStretch}{1.05}
\usepackage{textcomp}
\usepackage{xcolor}
\def\BibTeX{{\rm B\kern-.05em{\sc i\kern-.025em b}\kern-.08em
    T\kern-.1667em\lower.7ex\hbox{E}\kern-.125emX}}

\usepackage{acronym}
\usepackage{url}

\usepackage{lipsum}
\usepackage{fancyhdr}
\fancypagestyle{sec}{\lhead{\tmpx}}

\fancypagestyle{arXiv}
{
   \fancyhf{}
   \chead{\textcolor{gray}{This paper has been accepted to the 2026 IEEE Conference on Global Communications (GLOBECOM): \\
   Selected Areas in Communications: Machine Learning for Communications and Networking}}
   \cfoot{ \fontsize{=9pt}{10pt}\selectfont  \textcolor{gray}{© 2026 IEEE. Personal use of this material is permitted. Permission from IEEE must be obtained for all other uses, in any current or future media, including reprinting/republishing this material for advertising or promotional purposes, creating new collective works, for resale or redistribution to servers or lists, or reuse of any copyrighted component of this work in other works}\\\fontsize{=10pt}{12pt}\selectfont\thepage}
   
}

\newacro{RF}{Radio Frequency}
\newacro{PAs}{Power Amplifiers}
\newacro{PA}{Power Amplifier}
\newacro{DPD}{Digital Pre-distortion}
\newacro{GMP}{Generalized Memory Polynomial}
\newacro{NNs}{Neural Networks}
\newacro{NN}{Neural Network}
\newacro{DNNs}{deep neural networks}
\newacro{ML}{Machine Learning}
\newacro{RNN}{Recurrent Neural Network}
\newacro{RNNs}{Recurrent Neural Networks}
\newacro{OFDM}{Orthogonal Frequency Division Modulation}
\newacro{ADC}{Analog-to-Digital Converters}
\newacro{EVM}{Error Vector Magnitude}
\newacro{ACPR}{Adjacent Channel Power Ratio}
\newacro{NMSE}{Normalized Mean Square Error}
\newacro{MSE}{Mean Square Error}
\newacro{CNN}{Convolutional Neural Network}
\newacro{GRU}{Gated Recurrent Unit}
\newacro{TCN}{Temporal Convolutional Networks}
\newacro{MP}{Mixed Precision}
\newacro{LSTM}{Long Short-Term Memory}
\newacro{JANET}{Just Another NETwork}
\newacro{RRU}{Residual Recurrent Unit}
\newacro{PN}{Phase Normalization}
\newacro{BO}{Block-oriented}
\newacro{FC}{Fully-Connected}
\newacro{PAPR}{Peak-to-Average Power Ratio}
\newacro{DVR}{Decomposed Vector Rotation}
\newacro{DGRU}{Dense Gated Recurrent Unit}
\newacro{PG}{Phase Gated}
\newacro{TRes-DeltaGRU}{Temporal Residual Delta Gated Recurrent Unit}
\newacro{DeltaGRU}{Delta Gated Recurrent Unit}
\newacro{FCRes-DeltaGRU}{Fully Connected Residual Delta Gated Recurrent Unit}
\newacro{TRes}{temporal residual}
\newacro{TOR}{Transmitter Observation Receiver}
\newacro{ISA}{Instruction Set Architecture}
\newacro{BPTT}{BackPropagation Through Time}
\newacro{ASIC}{Application-Specific Integrated Circuit}
\newacro{CCDF}{Complementary Cumulative Distribution Function}
\newacro{FLOPs}{Floating-Point Operations}
\newacro{APNRRU}{Augmented Phase Normalization Residual Recurrent Unit}

\begin{document}
\title{OpenDPDv2: A Unified Learning and Optimization Framework for Neural Network Digital Predistortion}

% \author{\IEEEauthorblockN{Yizhuo Wu}
% \IEEEauthorblockA{\textit{Department of Microelectronics} \\
% \textit{Delft University of Technology}\\
% Delft, the Netherlands \\
% yizhuo.wu@tudelft.nl}
% \and
% \IEEEauthorblockN{Ang Li}
% \IEEEauthorblockA{\textit{Department of Microelectronics} \\
% \textit{Delft University of Technology}\\
% Delft, the Netherlands \\
% ang.li@tudelft.nl}
% \and
% \IEEEauthorblockN{Chang Gao}
% \IEEEauthorblockA{\textit{Department of Microelectronics} \\
% \textit{Delft University of Technology}\\
% Delft, the Netherlands \\
% chang.gao@tudelft.nl}
% }

\author{
Yizhuo Wu,
Ang Li, and Chang Gao*\\
Department of Microelectronics, Delft University of Technology, Delft 2628 CD, The Netherlands \\
Email: \{yizhuo.wu, ang.li, chang.gao\}@tudelft.nl
\thanks{*Corresponding author: Chang Gao}
}
\IEEEaftertitletext{\vspace{-1.0\baselineskip}}

\maketitle

\begin{abstract}
Neural network (NN)-based Digital Predistortion (DPD) improves linearization for wideband radio frequency (RF) power amplifiers (PAs) but often increases the complexity of the digital back-end. This paper presents \texttt{OpenDPDv2}, an open-source end-to-end framework that unifies PA modeling, NN-DPD learning, and deployment-oriented optimization. \texttt{OpenDPDv2} introduces \ac{TRes}-\ac{DeltaGRU}, a delta-RNN DPD architecture with a lightweight temporal residual path for robust operation under aggressive temporal sparsity, and supports joint optimization with fixed-point quantization. On a 3.5\,GHz GaN Doherty PA driven by a TM3.1a 200\,MHz 256-QAM OFDM signal, the FP32 \ac{TRes}-\ac{DeltaGRU} model achieves $-59.9$\,dBc \ac{ACPR} and $-42.1$\,dB \ac{EVM}. With 56\% temporal sparsity and W12A12 quantization, the model uses 450 active parameters while maintaining $-51.8$\,dBc \ac{ACPR} and $-35.2$\,dB \ac{EVM}. Timing-accurate \texttt{Gem5}-based evaluation further indicates a 4.5$\times$ reduction in forward-pass energy for this deployment-oriented operating point. Code, datasets, and documentation are publicly available at \url{https://github.com/lab-emi/OpenDPD}.
\end{abstract}


\section{Introduction}
\thispagestyle{arXiv}
The exponential growth of wireless data traffic has increased the demand for wider spectrum use, higher data rates, and lower error rates in modern communication systems. However, wideband \ac{RF} \ac{PAs}, a key component of transmitters, introduce inherent distortions that increase out-of-band emissions and degrade in-band signal quality, resulting in higher \ac{EVM} and symbol errors, thereby reducing both communication reliability and energy efficiency. To mitigate these effects, \ac{DPD} is widely adopted. 

\ac{DPD} seeks to approximate the inverse transfer function of the \ac{PA} to pre-compensate nonlinear distortion before transmission. Conventional Volterra-series-based \ac{DPD} methods provide effective linearization in narrowband scenarios, but often fail to meet communication standards for wideband, high-order modulated signals~\cite{3GPP}. In contrast, \ac{ML} methods, particularly \ac{NN}-based \ac{DPD}, have been shown to outperform the \ac{GMP} model for \ac{OFDM} signals above 200\,MHz while satisfying standard requirements~\cite{PGJANET,DVRJANET,PNRNN,BOJANET}. 

However, the \ac{DPD} block itself consumes a significant share of the power in wideband digital radio back-ends~\cite{wesemann2023energy}, and the use of \ac{NNs} can further aggravate this issue. Such overhead conflicts with the goal of energy-efficient transmitters in base stations and Wi-Fi access points, where power budgets are constrained. 

Prior efforts to reduce \ac{DPD} energy consumption include lowering the sample rate~\cite{Li2020SampleRate}, using a sub-Nyquist feedback receiver in the observation path~\cite{Hammler2019}, dynamically adapting model cross-terms to the input signal~\cite{Li2022}, simplifying DPD computational paths~\cite{Beikmirza2023}, and pruning less important weights in fully connected layers to induce static spatial sparsity~\cite{Liu2022}. 

This work, \texttt{OpenDPDv2}, extends the end-to-end \ac{DPD} learning of \texttt{OpenDPDv1} and the threshold-based recurrent sparsity of DeltaDPD~\cite{wu2024opendpd,WuIMS2025,Wu2024IMS} by introducing \ac{TRes}-\ac{DeltaGRU} to decouple output dynamics from hidden-state sparsity, jointly optimizing sparsity and quantization, and evaluating processor-level energy with timing-accurate \texttt{Gem5}.

The main contributions are: 

\begin{enumerate} 
\item \textbf{TRes-DeltaGRU: a delta-RNN DPD architecture that decouples output dynamics from hidden-state sparsity.} We propose \ac{TRes}-\ac{DeltaGRU}, which adds a \ac{TCN}-based temporal residual path to improve robustness under aggressive sparsity. For a 200\,MHz TM3.1a 256-QAM \ac{OFDM} signal on a 3.5\,GHz GaN Doherty \ac{PA} at 41.2\,dBm average output power, \ac{TRes}-\ac{DeltaGRU} achieves 72.5\% sparsity with 288 active parameters while delivering measured performance of -52.0\,dBc \ac{ACPR} and -37.0\,dB \ac{EVM}. 

\item \textbf{\texttt{OpenDPDv2}: a reproducible end-to-end learning and optimization pipeline.} We release \texttt{OpenDPDv2}, a unified \texttt{PyTorch}-based framework that (i) trains \ac{DPD} models end-to-end through a frozen \ac{PA} surrogate and (ii) supports joint optimization with delta-threshold temporal sparsity and quantization-aware training. For the TM3.1a 200\,MHz test signal, \texttt{OpenDPDv2} supports \ac{TRes}-\ac{DeltaGRU} variants down to 450 active parameters at 56.0\% temporal sparsity while maintaining -51.8\,dBc \ac{ACPR} and -35.2\,dB \ac{EVM} with W12A12 quantization. 

\begin{figure*}
    \centering
    \includegraphics[width=\linewidth]{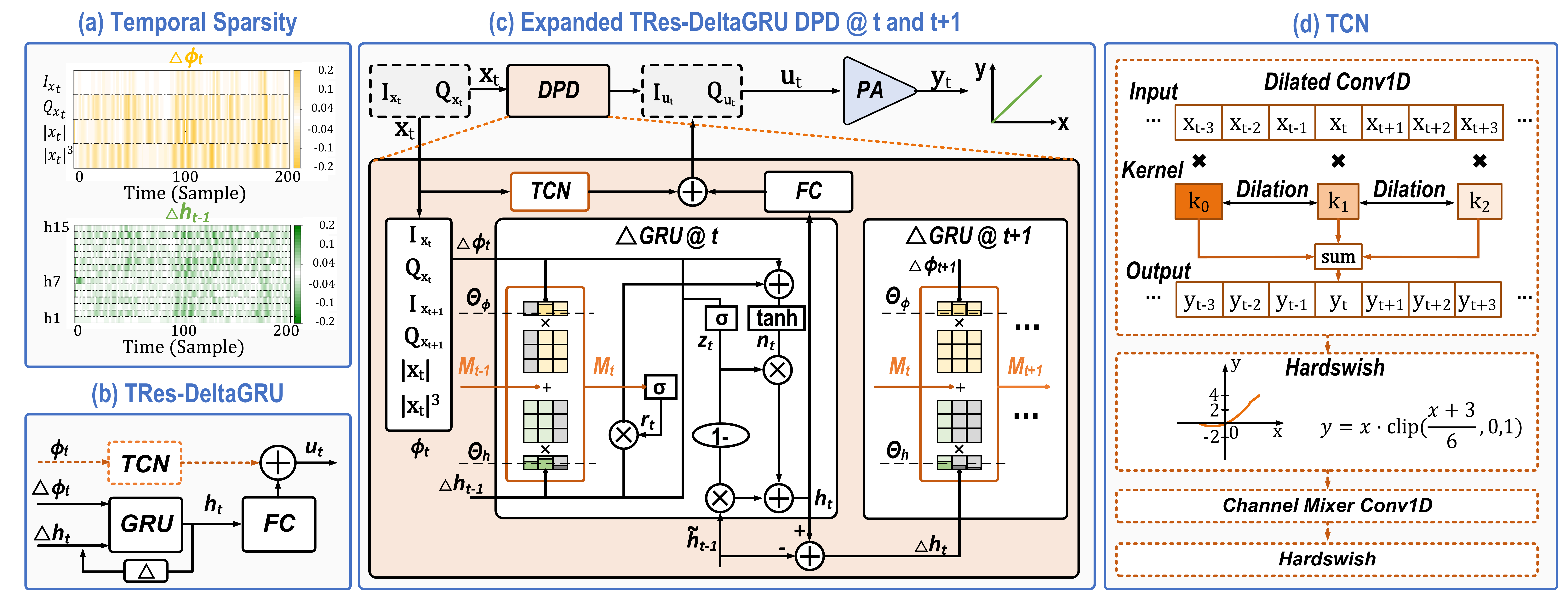}
    \caption{(a) Dynamic temporal sparsity in recurrent neural network-based DPD input and hidden neurons. (b) TRes-DeltaGRU diagram. (c) Unfolded TRes-DeltaGRU cell in the DPD model with temporal sparsity. (d) Architecture of the \ac{TCN}-based residual path.}
    \label{fig:method}
\end{figure*}

\item \textbf{Gem5-based deployment analysis for power-aware DPD optimization.} To complement measured linearization results, we evaluate optimized \ac{TRes}-\ac{DeltaGRU} variants on a timing-accurate \texttt{Gem5} model of a 32-bit ARMv7-A embedded processor. For the 999-parameter model, W12A12 quantization alone reduces forward-pass energy by 2.8$\times$, and combining it with 56.0\%/72.5\% temporal sparsity yields 4.5$\times$/5.2$\times$ reduction while still meeting a common -45\,dBc \ac{ACPR} target. 
\end{enumerate}
%--------------------------------------
\section{Proposed TRes-DeltaGRU}
%--------------------------------------

%----------------------------------------
\subsection{Dynamic Temporal Sparsity in DPD Signals and Hidden States}
\label{seq:DeltaDPD}
%----------------------------------------

The delta-processing principle is motivated by the temporal redundancy of \ac{DPD} signals. Considering the input feature set $\begin{bmatrix} I_{x_t},\,Q_{x_t},\,|x_t|,\,|x_t|^3 \end{bmatrix}$ and a 15-neuron hidden-state \ac{GRU}-based \ac{DPD} model as an example, Fig.~\ref{fig:method}(a) visualizes the element-wise changes between adjacent time steps in the input neurons $\Delta\phi$ and hidden states $\Delta h$. White pixels indicate Delta values close to zero, showing that both the input features and hidden states vary slowly over time. This temporal redundancy enables dense-matrix-dense-vector multiplication (\textbf{M$\times$V}) to be reformulated as dense-matrix-sparse-vector multiplication (\textbf{M$\times$SV}).

%--------------------------------------
\subsection{TRes-DeltaGRU Algorithm}
%--------------------------------------
In the Delta network algorithm, the sequential \textbf{M$\times$V} operation is rewritten as an incremental update driven by temporal differences:
\begin{align}
\mathbf{y}_t &= \mathbf{W} \mathbf{v}_t, \label{eq1} \\
\mathbf{y}_t &= \mathbf{W} \Delta\mathbf{v}_t + \mathbf{y}_{t-1}
            = \mathbf{W}(\mathbf{v}_t - \mathbf{v}_{t-1}) + \mathbf{y}_{t-1}, \label{eq2}
\end{align}
where $\mathbf{v}_t$ denotes either the \ac{RNN} input $\boldsymbol{\phi}_t = \begin{bmatrix} I_t,\,Q_t,\,I_{t+1}, Q_{t+1},|x_t|,\,|x_t|^3 \end{bmatrix}$ or the hidden state $\mathbf{h}_{t-1}$. We sparsify $\Delta\mathbf{v}_t$ with a threshold $\Theta_v$, so that only meaningful changes are propagated:
\begin{align}
\Delta\mathbf{v}_t =
\begin{cases}
\mathbf{v}_t - \tilde{\mathbf{v}}_{t-1}, & |\mathbf{v}_t - \tilde{\mathbf{v}}_{t-1}| > \Theta_{v}, \\
0, & |\mathbf{v}_t - \tilde{\mathbf{v}}_{t-1}| \leq \Theta_{v}, \label{eq4}
\end{cases}
\end{align}
where $\tilde{\mathbf{v}}$ stores the latest effective value. For the $k$-th element, the buffer is updated only when the change exceeds the threshold:
\begin{align}
\tilde{v}^{k}_{t-1} =
\begin{cases}
v^{k}_{t-1}, & |v^{k}_t - \tilde{v}^{k}_{t-1}| > \Theta_{v}, \\
\tilde{v}^{k}_{t-2}, & |v^{k}_t - \tilde{v}^{k}_{t-1}| \leq \Theta_{v}. 
\label{eq5}
\end{cases}
\end{align}

The delta formulation of the \ac{GRU} update is given by:
\begin{align}
\mathbf{r}_t &= \sigma\left(\mathbf{M}_{r,t}\right), \label{eq10}\\
\mathbf{z}_t &= \sigma\left(\mathbf{M}_{z,t}\right), \\
\mathbf{n}_t &= \tanh\left(\mathbf{M}_{n\phi,t} + \mathbf{r}_t \odot \mathbf{M}_{nh,t}\right), \\
\mathbf{h}_t &= \left(1 - \mathbf{z}_t\right) \odot \mathbf{h}_{t-1} + \mathbf{z}_t \odot \mathbf{n}_t. \label{eq13}
\end{align}
The pre-activation accumulators $\mathbf{M}_{r,t}$, $\mathbf{M}_{z,t}$, $\mathbf{M}_{n\phi,t}$, and $\mathbf{M}_{nh,t}$ are initialized by $\mathbf{M}_{r,0} = \mathbf{b}_{ir} + \mathbf{b}_{hr}$, $\mathbf{M}_{z,0} = \mathbf{b}_{iz} + \mathbf{b}_{hz}$, $\mathbf{M}_{n\phi,0} = \mathbf{b}_{in}$, and $\mathbf{M}_{nh,0} = \mathbf{b}_{hn}$, and updated incrementally as:
\begin{align}
\mathbf{M}_{r,t}     &= \mathbf{W}_{ir}\Delta\boldsymbol{\phi}_t + \mathbf{W}_{hr}\Delta\mathbf{h}_{t-1} + \mathbf{M}_{r,t-1}, \label{eq15}\\
\mathbf{M}_{z,t}     &= \mathbf{W}_{iz}\Delta\boldsymbol{\phi}_t + \mathbf{W}_{hz}\Delta\mathbf{h}_{t-1} + \mathbf{M}_{z,t-1}, \label{eq16}\\
\mathbf{M}_{n\phi,t} &= \mathbf{W}_{in}\Delta\boldsymbol{\phi}_t + \mathbf{M}_{n\phi,t-1}, \label{eq17}\\
\mathbf{M}_{nh,t}    &= \mathbf{W}_{hn}\Delta\mathbf{h}_{t-1} + \mathbf{M}_{nh,t-1}. \label{eq18}
\end{align}

In a plain DeltaGRU-based DPD~\cite{WuIMS2025}, the final output is generated directly from the recurrent hidden state through a linear layer. This couples the output dynamics to hidden-state sparsity and can overly constrain the predistorted waveform under aggressive delta thresholds. For example, at $\Theta_h=0.05$, the relative sparsity of $\mathbf{h}_t$ reaches 81.8\%, whereas the corresponding output waveform exhibits only 42.4\% relative sparsity. This gap indicates that enforcing sparsity only in the recurrent state can suppress rapid output variations that are still necessary for accurate linearization.

Fig.~\ref{fig:method}(b) summarizes \ac{TRes-DeltaGRU}: a DeltaGRU branch is combined with a lightweight \ac{TCN} input-to-output residual path. Fig.~\ref{fig:method}(c) unfolds the sparse recurrent branch over time, with gray elements marking skipped computations. By bypassing the recurrent state, the \ac{TCN} preserves fast local waveform variations and decouples output dynamics from hidden-state sparsity.

The \ac{TCN} uses dilated convolutions, which expand the temporal receptive field from $K$ to $(K-1)\times D + 1$ without increasing the parameter count, as illustrated in Fig.~\ref{fig:method}(d). The residual path is defined as
\begin{align}
    \text{TCN}= [\text{Dilated Conv1D},\text{Hardswish},\text{Conv1D},\text{Hardswish}]
\end{align}
where Hardswish is defined as
\begin{align}
    y=x\cdot \text{clip}(\frac{x+3}{6},0,1).
\end{align}
Hardswish is used because it keeps a smooth gradient around zero for stable training while remaining hardware-friendly for efficient fixed-point implementation.

The final \ac{DPD} output, $\mathbf{U}$, is expressed as
\begin{align}
\begin{bmatrix} \mathbf{I_u},\mathbf{Q_u}\end{bmatrix}=\mathbf{U} &= \mathbf{\hat{U}} +\text{TCN} (\mathbf{X})
\end{align}
where $\mathbf{\hat{U}} = \{\mathbf{\hat{u}}_t \mid \mathbf{\hat{u}}_t = \mathbf{W}_{\hat{y}}\boldsymbol{h}_{t} + \mathbf{b}_{\hat{y}}, \, \boldsymbol{h}_{t}\in \mathbb{R}^H, t \in 0,\dots ,T-1\}$ and $H$ is the hidden size. Delta thresholds $\Theta_{\phi}$ and $\Theta_{h}$ are applied to skip \textbf{MAC} operations associated with small $\Delta$ entries and the corresponding weight-column accesses.

%----------------------------------------
\section{OpenDPDv2}
\label{seq:OpenDPDv2}
%----------------------------------------
\begin{figure}
    \centering
    \includegraphics[width=0.95\linewidth]{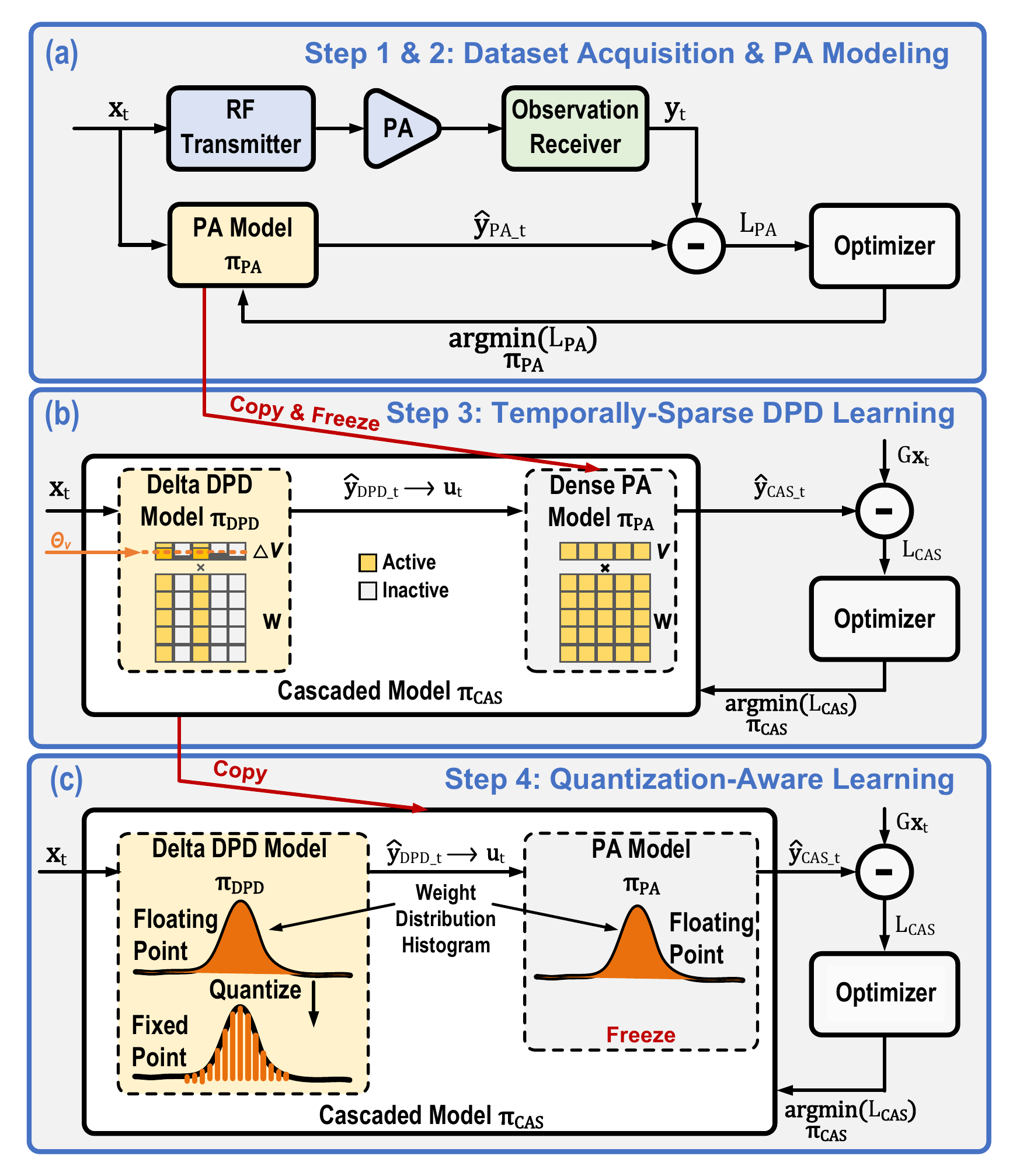}
    \caption{Learning and optimization flow of \texttt{OpenDPDv2}.}
    \label{fig:opendpdv2}
\end{figure}

To train and optimize the proposed model, we use \texttt{OpenDPDv2}, a unified end-to-end learning and deployment-oriented optimization flow shown in Fig.~\ref{fig:opendpdv2}. End-to-end training is realized by cascading the \ac{DPD} model with a pre-trained, frozen \ac{PA} surrogate and optimizing the \ac{DPD} parameters using a loss defined at the \ac{PA} output. In this way, the learned \ac{DPD} directly compensates the nonlinear \ac{PA} behavior in the cascaded system. This unified flow also makes the comparison between dense and sparse models more consistent, since the same training objective, \ac{PA} surrogate, and evaluation procedure are maintained throughout the optimization process. It therefore provides a practical basis for studying how temporal sparsity and fixed-point quantization affect the linearization-efficiency trade-off within a single reproducible pipeline.

\texttt{OpenDPDv2} contains three stages:
\begin{itemize}
\item \textbf{Data acquisition and PA modeling:} collect the baseband input/output pair $(\mathbf{X},\mathbf{Y})$ and train a behavioral \ac{PA} model $\pi_{PA}$ with \ac{BPTT}.
\item \textbf{Temporally sparse DPD learning:} place \ac{TRes-DeltaGRU} before the frozen $\pi_{PA}$ and train the cascade to map $\mathbf{x}_t$ to $G\mathbf{x}_t$, so that the learned \ac{DPD} approximates the inverse \ac{PA} response.
\item \textbf{Quantization-aware optimization:} apply quantization-aware training~\cite{Nagel2021} using quantized weights and activations in the forward pass while updating full-precision variables during backpropagation.
\end{itemize}

Selected sparse and quantized operating points are exported to \texttt{C} and evaluated with \texttt{Gem5} to connect algorithm-level compression with processor-level energy behavior.

%-----------------------------------------------------------------------------
\section{Experimental Setup}
%-----------------------------------------------------------------------------

\subsection{Measurement Setup}
\begin{figure}[!t]
    \centering
    \includegraphics[width=0.9\linewidth]{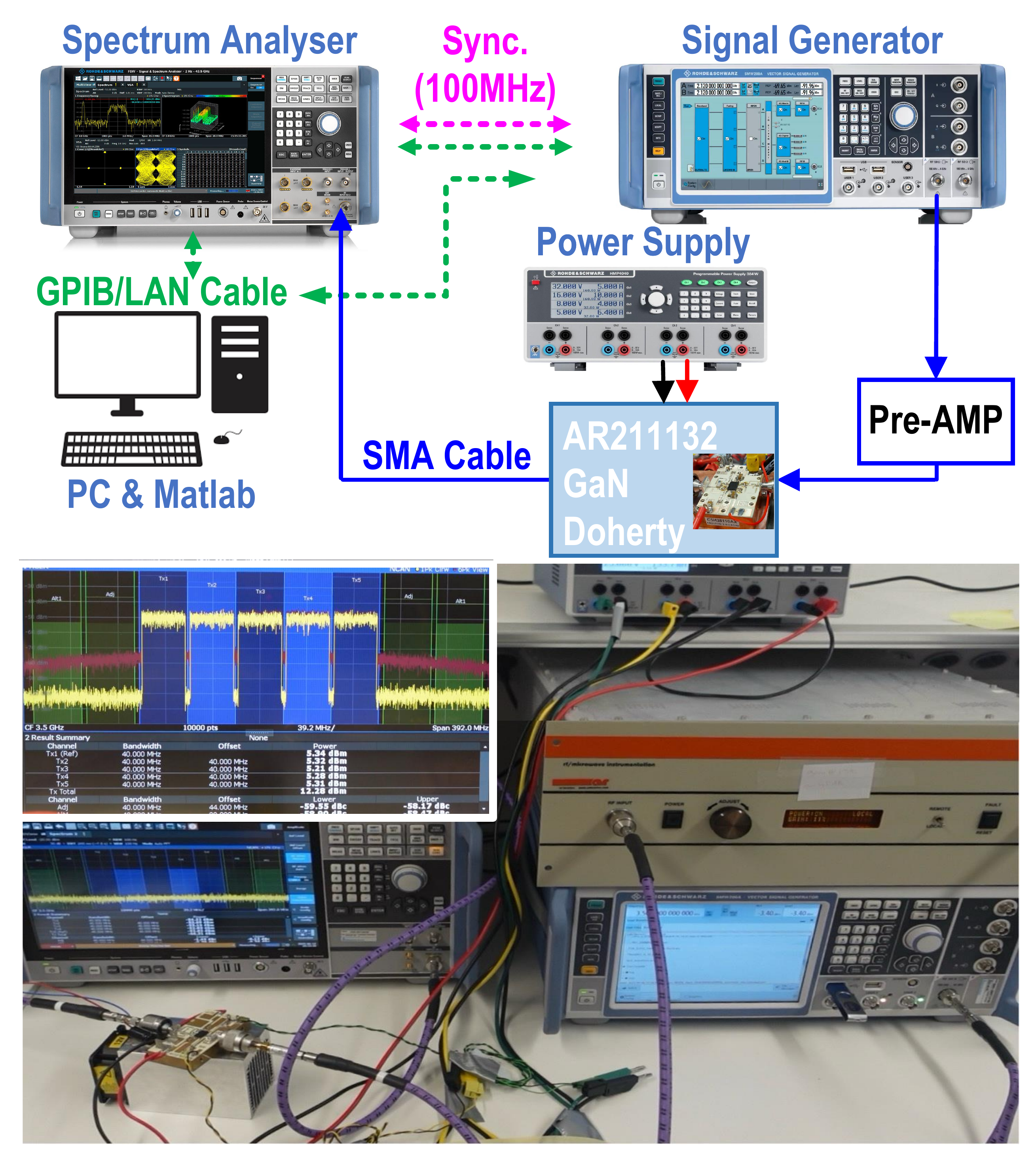}
    \caption{Setup for dataset acquisition and DPD performance measurement.}
    \label{fig:platform}
\end{figure}

Figure~\ref{fig:platform} shows the experimental setup used in this study. The \ac{OFDM} baseband I/Q signal was generated by an \texttt{R\&S-SMW200A} signal generator and amplified by the \ac{PA} at the same average output power, with and without \ac{DPD}. The output signal was then digitized by an \texttt{R\&S-FSW43} spectrum analyzer. For \ac{ACPR} measurements, the resolution bandwidth was set to 100\,Hz to lower the analyzer noise floor~\cite{SA_RBW}. The \ac{EVM} was computed by comparing the input signal with the digitized output signal rather than the reference grid.
\subsection{Test Signals and Design-Under-Test (DUT)}
The experiment used \texttt{APA\_200MHz}, a \texttt{TM3.1a} 5$\times$40-MHz (200-MHz) 256-QAM signal with a \ac{PAPR} of 10.01\,dB at \ac{CCDF} = 0.001\% (9.24\,dB at \ac{CCDF} = 0.01\%) and a sampling rate of 983.04\,MHz. The dataset contained 98\,304 samples, split into 60\%/20\%/20\% for training, validation, and testing. Experiments were conducted on a 3.5\,GHz GaN Doherty \ac{PA} DUT (internal evaluation board AR211132 from Ampleon), operating at an average output power of 41.2\,dBm. The \ac{PA} exhibits a P1dB of 46.5\,dBm, a P3dB of 50\,dBm, and a flat gain bandwidth of 200\,MHz.

\subsection{Neural Network Training}
End-to-end training was implemented on \texttt{OpenDPDv2}. The pre-trained 2751-parameter DGRU PA behavioral model achieved -39.96\,dB-NMSE. Models were trained offline for 240 epochs using ADAMW with an initial learning rate of 5E-3, a \texttt{ReduceOnPlateau} schedule, and a batch size of 64. In the model comparison experiment, the \ac{DVR} decomposition used 3 units, the \ac{PG}-\ac{JANET} model used a window size of 4, both \ac{JANET} models used a hidden size of 12, and \ac{TRes-DeltaGRU} used 15 hidden neurons. Each Conv1D tuple denotes m$(C_{in},C_{out},K,D,P)$: input/output channels, kernel size, dilation, and padding. Thus, the dilated layer configuration is $(2,3,3,16,16)$, while the channel mixer configuration is $(3,2,1,1,1)$. All experiments were run on a single NVIDIA RTX 4090 GPU with PyTorch 2.4.1 and CUDA 12.4, and the checkpoint with the best \ac{ACPR} was used for final evaluation.

\subsection{Gem5-Based Energy Estimation Setup}
To complement the measured linearization results, forward-pass energy was evaluated using the timing-accurate \texttt{Gem5} simulator~\cite{gem5}. The DPD inference kernels were implemented in \texttt{C} and cross-compiled for a 32-bit ARMv7-A processor using \texttt{arm-linux-gnueabihf-gcc} with \texttt{-O3 -march=armv7-a -mfpu=neon -static}. The simulated single-core processor uses separate 8\,KB 4-way L1 instruction and data caches and 1\,GB DDR4 main memory, and energy is estimated using 7\,nm operation and memory numbers~\cite{jouppi2021ten}. Each run processes an I/Q sequence of 10\,000 samples. Since the entire DPD model fits in the 8\,KB D-cache, DDR4 energy is excluded after initialization, and nearly 100\% L1 data-cache hit rate is observed.

\section{Results and Discussion}

\begin{table}[!t]
    \centering
    \caption{Comparison of neural DPD models trained by OpenDPDv2}
    \label{tab:model}
    \TblFont
    \setlength{\tabcolsep}{\TblColSep}
    \renewcommand{\arraystretch}{\TblStretch}
    \scalebox{\TblScale}{%
    \begin{threeparttable}
    \begin{tabular}{|c|c|c|ccc|}
    \hline
    \textbf{\begin{tabular}[c]{@{}c@{}}Dense \\ DPD Models\end{tabular}} & \textbf{\#Params} & \textbf{FLOPs\tnote{a}} & \textbf{\begin{tabular}[c]{@{}c@{}}NMSE\\ (dB)\end{tabular}} & \textbf{\begin{tabular}[c]{@{}c@{}}EVM\\ (dB)\end{tabular}} & \textbf{\begin{tabular}[c]{@{}c@{}}ACPR\\ (dBc)\end{tabular}} \\ \hline\hline
    Without DPD & - & - & -20.5 & -24.7 (5.84\%) & -28.3 \\ \hline\hline
    RVTDCNN~\cite{Hu2022RVTDCNN} & 1007 & 1587 (+23.8\%) & -32.9 & -33.9 (2.03\%) & -50.8 \\
    PG-JANET~\cite{PGJANET} & 1130  & 1507 (+17.6\%) & -36.6 & -39.3 (1.09\%) & -58.2 \\
    DVR-JANET~\cite{DVRJANET} & 1097 & 1370 (+6.9\%) & -37.0 & -38.6 (1.17\%) & -59.4 \\
    BO-JANET~\cite{BOJANET} & 1064 & 1535 (+19.7\%) & -39.8 & -42.9 (0.71\%) & -58.7 \\
    APNRRU~\cite{APNRRU} & 1043 & 1328 (+3.5\%) & -36.0 & -38.1 (1.25\%) & -58.6 \\
     \textbf{TRes-GRU} &  \textbf{999} &  \textbf{1282} &  \textbf{-38.4} &  \textbf{-41.2 (0.87\%)} &  \textbf{-59.0} \\
     \textbf{TRes-DeltaGRU} &  \textbf{999} &  \textbf{1324 (+3.3\%)} &  \textbf{-39.6} &  \textbf{-42.1 (0.79\%)} & \textbf{-59.9} \\\hline
    \end{tabular}%
    \begin{tablenotes}
    \item[a] Percentage values indicate the difference relative to TRes-GRU.
    \end{tablenotes}
    \end{threeparttable}}
    
\end{table}

We compare the absolute linearization performance of \texttt{OpenDPDv2}-trained TRes-DeltaGRU DPD models of different complexities against prior work. For an intuitive measure of sparse-model complexity, the number of active parameters in TRes-DeltaGRU is defined as
\begin{equation}
\begin{aligned}
\text{\#}\text{Active Params} &= \text{\#}\text{DeltaGRU Params} \times (1-\Gamma) \\
&\quad + \text{\#}\text{FC Params} + \text{\#}\text{TCN Params}
\end{aligned}
\end{equation}
where $\Gamma$ denotes the temporal sparsity. In the delta formulation, zero entries in $\Delta\boldsymbol{\phi}$ and $\Delta\mathbf{h}$ skip the corresponding \textbf{MAC} operations and associated weight-column accesses; therefore, the number of active parameters serves as an implementation-oriented proxy for both active computation and weight-access activity.

\subsection{Model Comparison}
Table~\ref{tab:model} compares \ac{TRes}-\ac{DeltaGRU} with five prior \ac{DPD} models of about 1000 parameters. It achieves the best overall balance, reaching -39.6\,dB \ac{NMSE}, -42.1\,dB (0.79\%) \ac{EVM}, and -59.9\,dBc \ac{ACPR}. Among the baselines, \ac{BO}-\ac{JANET} provides the best \ac{EVM} and \ac{DVR}-\ac{JANET} the best \ac{ACPR}, but \ac{TRes}-\ac{DeltaGRU} improves on both simultaneously. Quantitatively, it improves \ac{ACPR} over \ac{BO}-\ac{JANET} by 1.2\,dB and improves \ac{EVM} over \ac{DVR}-\ac{JANET} by 3.5\,dB. From an efficiency perspective, it is also the smallest model in the comparison (999 parameters), and its dense FLOPs remain close to those of \ac{TRes}-\ac{GRU}, with only 3.3\% overhead. This makes \ac{TRes}-\ac{DeltaGRU} a suitable dense starting point for subsequent sparsity and quantization optimization.

\subsection{Ablation Study}

\begin{table}[!t]
    \caption{Ablation study of temporally sparse DPD models}
    \label{tab:ablation}
    \centering
    {\TblFont
    \setlength{\tabcolsep}{\TblColSep}
    \renewcommand{\arraystretch}{\TblStretch}
    \begin{tabular}{|c|c|cc|ccc|}
    \hline
    \textbf{DPD Model} & \multicolumn{1}{c|}{\textbf{Temporal}} & \multicolumn{2}{c|}{\textbf{Params}} & \multicolumn{1}{c}{\textbf{NMSE}} & \textbf{EVM} & \textbf{ACPR} \\
    \textbf{Type} & \multicolumn{1}{c|}{\textbf{Sparsity}} & \multicolumn{1}{l}{\textbf{\#Total}} & \multicolumn{1}{l|}{\textbf{\#Active}} & \multicolumn{1}{c}{\textbf{(dB)}} & \textbf{(dB)} & \textbf{(dBc)} \\ \hline \hline
    \multirow{3}{*}{\textbf{DeltaGRU}} & 0\% & \multirow{3}{*}{1090} & 1090 & -37.1 & -38.6 (1.17\%) & -59.4 \\
     & 56.0\% &  & 498 & -34.9 & -37.0 (1.41\%) & -52.8 \\
     & 75.8\% &  & 290 & -34.6 & -36.7 (1.47\%) & -48.4 \\ \hline
    \multirow{3}{*}{\textbf{\begin{tabular}[c]{@{}c@{}}FCRes\\ -DeltaGRU\end{tabular}}} & 0\% & \multirow{3}{*}{1094} & 1094 & -37.1 & -38.4 (1.20\%) & -58.0 \\
     & 56.2\% &  & 501 & -36.1 & -38.2 (1.23\%) & -51.4 \\
     & 76.1\% &  & 290 & -35.6 & -37.0 (1.41\%) & -50.1 \\ \hline
    \multirow{3}{*}{\textbf{\begin{tabular}[c]{@{}c@{}}TRes\\ -DeltaGRU\end{tabular}}} & 0\% & \multirow{3}{*}{999} & 999 & -39.8 & -42.1 (0.79\%) & -59.9 \\
     & 56.0\% &  & 450 & -38.1 & -40.6 (0.93\%) & -52.9 \\
     & 72.5\% &  & 288 & -36.0 & -37.0 (1.41\%) & -52.0 \\ \hline
    \end{tabular}%
    }%
\end{table}
Table~\ref{tab:ablation} compares three temporally sparse models: \ac{DeltaGRU} without a residual connection, \ac{FCRes-DeltaGRU} with a linear residual path, and \ac{TRes-DeltaGRU} with a \ac{TCN} residual path. The residual connection clearly mitigates performance loss under temporal sparsity, and the \ac{TCN}-based version is best across all operating points. At 0\% sparsity, \ac{TRes-DeltaGRU} already achieves the best performance. Around 56\% sparsity, plain \ac{DeltaGRU} shows 3.6\,dB worse \ac{EVM} than \ac{TRes-DeltaGRU} while using more active parameters; at about 300 active parameters, \ac{TRes-DeltaGRU} still achieves -52.0\,dBc \ac{ACPR}. These results verify the effectiveness of the \ac{TCN} residual path in preserving output fidelity at high sparsity.

\subsection{Sparsity and Quantization Trade-off}
\begin{table}[!t]
    \caption{Linearization performance of \ac{TRes}-\ac{GRU} (Dense) vs. TRes-DeltaGRU (Sparse) models.}
    \label{tab:optimization}
    \centering
    {\TblFont
    \setlength{\tabcolsep}{\TblColSep}
    \renewcommand{\arraystretch}{\TblStretch}
    \scalebox{\TblScale}{%
    \begin{tabular}{|ccccccc|}
    \hline
    \textbf{Model} & \textbf{Temporal} & \multicolumn{2}{c}{\textbf{Params}} & \textbf{Model} & \textbf{EVM} & \textbf{ACPR}\\
    \textbf{Type} & \textbf{Sparsity} & \multicolumn{1}{l}{\textbf{\#Total}} & \multicolumn{1}{l}{\textbf{\#Active}} & \textbf{Precision} & \textbf{(dB)} & \textbf{(dBc)} \\ \hline\hline
    \multicolumn{1}{|c|}{\textbf{Dense}} & - & 999 & 999 & FP32 & \textbf{-41.2 (0.87\%)} & -59.0 \\ \hline
    \multicolumn{1}{|c|}{\multirow{3}{*}{\textbf{Sparse}}} & \multirow{3}{*}{0\%} & \multirow{3}{*}{999} & \multirow{3}{*}{999} & FP32 & \textbf{-42.1 (0.79\%)} & -59.9 \\
    \multicolumn{1}{|c|}{} &  &  &  & W16A16 & -41.2 (0.87\%) & -58.8 \\
    \multicolumn{1}{|c|}{} &  &  &  & W12A12 & -37.3 (1.36\%) & -54.5 \\ \hline\hline
    \multicolumn{1}{|c|}{\textbf{Dense}} & - & 524 & 524 & FP32 & \textbf{-37.3 (1.36\%)} & -52.7 \\ \hline
    \multicolumn{1}{|c|}{\multirow{3}{*}{\textbf{Sparse}}} & \multirow{3}{*}{56.0\%} & \multirow{3}{*}{999} & \multirow{3}{*}{450} & FP32 & \textbf{-40.6 (0.93\%)} & -52.9 \\
    \multicolumn{1}{|c|}{} &  &  &  & W16A16 & -39.3 (1.08\%) & -53.2 \\
    \multicolumn{1}{|c|}{} &  &  &  & W12A12 & -35.2 (1.74\%) & -51.8 \\ \hline\hline
    \multicolumn{1}{|c|}{\textbf{Dense}} & - & 311 & 311 & FP32 & \textbf{-34.1 (1.97\%)} & -52.3 \\ \hline
    \multicolumn{1}{|c|}{\multirow{3}{*}{\textbf{Sparse}}} & \multirow{3}{*}{72.5\%} & \multirow{3}{*}{999} & \multirow{3}{*}{288} & FP32 & \textbf{-37.0 (1.41\%)} & -52.0 \\
    \multicolumn{1}{|c|}{} &  &  &  & W16A16 & -34.2 (1.95\%) & -48.2 \\
    \multicolumn{1}{|c|}{} &  &  &  & W12A12 & -31.0 (2.82\%) & -45.2 \\ \hline
    \end{tabular}%
    }%
    }
\end{table}
Table~\ref{tab:optimization} summarizes representative operating points from the joint threshold and precision scans. A key result is that sparse models can outperform dense ones at lower active complexity: W16A16 \ac{TRes-DeltaGRU}-450 improves on dense FP32 \ac{TRes}-\ac{GRU}-524 by 0.5\,dB in \ac{ACPR} and about 2\,dB in \ac{EVM} while using 14.1\% fewer active parameters, and FP32 \ac{TRes-DeltaGRU}-288 still achieves better \ac{EVM} than dense FP32 \ac{TRes}-\ac{GRU}-311. This shows that threshold-controlled temporal sparsity can improve the linearization-complexity trade-off rather than only degrade performance gracefully. From a deployment perspective, Table~\ref{tab:optimization} also exposes several useful operating regions rather than a single optimum. For example, the 56.0\% sparsity setting preserves strong linearization with only 450 active parameters, while the 72.5\% sparsity setting further reduces active complexity and remains compatible with the common -45\,dBc \ac{ACPR} target later used in the \texttt{Gem5}-based energy analysis.

For reference, 3GPP FR1 base-station limits are 3.5\% \ac{EVM} for 256-QAM ($-29.1$\,dB) and -45\,dB ACLR, equivalent to a $-45$\,dBc target under our signed \ac{ACPR} convention~\cite{3GPP}. FP32 \ac{TRes-DeltaGRU}-999 has 13.0/14.9\,dB \ac{EVM}/\ac{ACPR} margins to these limits. At 56.0\% sparsity, W12A12 \ac{TRes-DeltaGRU}-450 retains 6.1/6.8\,dB \ac{EVM}/\ac{ACPR} margins; at 72.5\%, W12A12 \ac{TRes-DeltaGRU}-288 retains 1.9/0.2\,dB. These points come from a sweep over $\Theta_{\phi}\in[0,0.04]$ and $\Theta_h\in[0,0.4]$; at the most aggressive setting, the FP32 model reaches 80.2\% sparsity and 222 active parameters, with -46.7\,dBc \ac{ACPR} and -35.1\,dB \ac{EVM}.

\subsection{Gem5-Based Energy Analysis}
\begin{figure}[!t]
    \centering
    \includegraphics[width=0.9\linewidth]{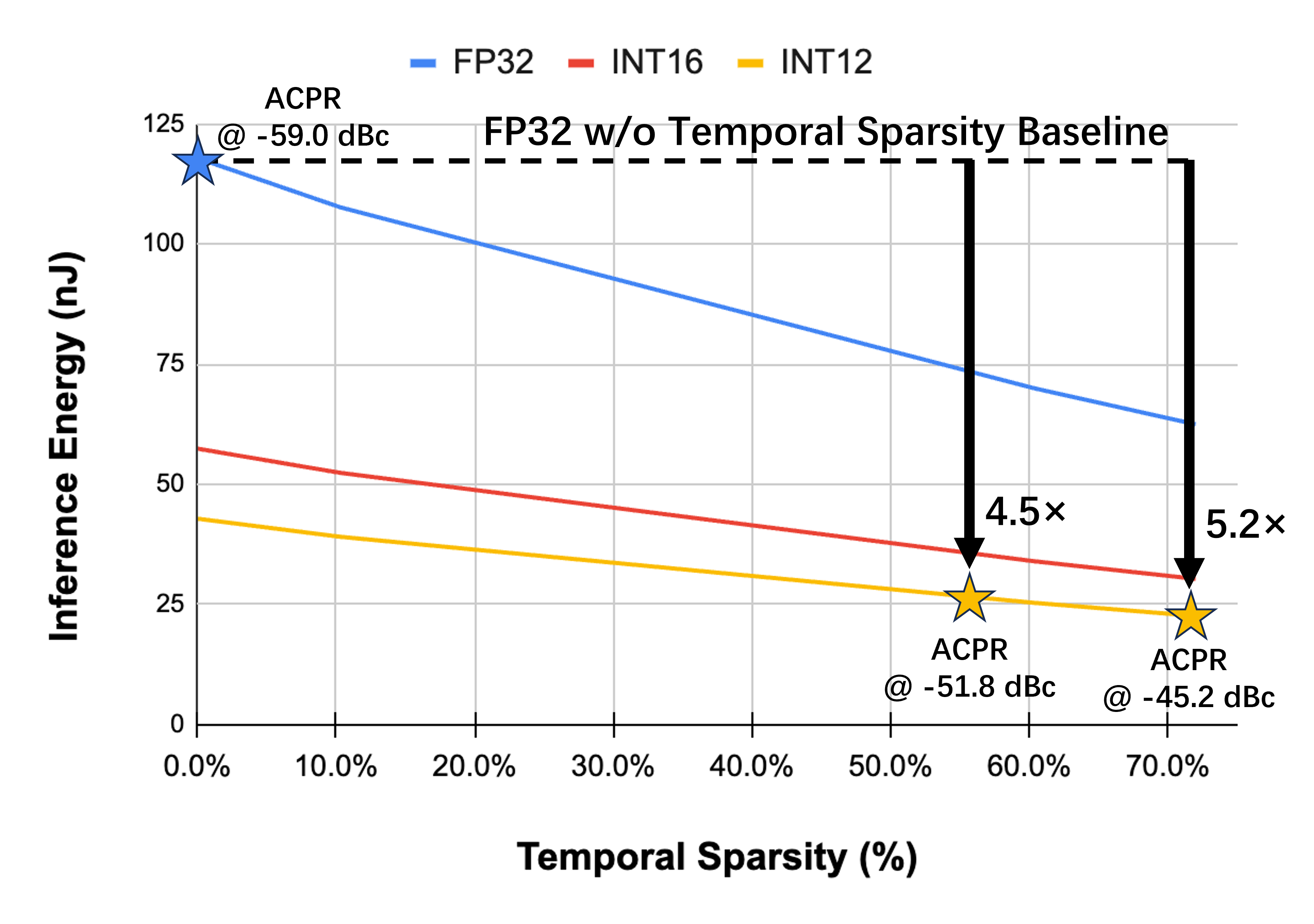}
    \caption{TRes-DeltaGRU-999 forward-pass energy reduction with temporal sparsity estimated using the \texttt{Gem5}-simulated ARM CPU.}
    \label{fig:energy_sparsity}
\end{figure}

Fig.~\ref{fig:energy_sparsity} shows the forward-pass energy of \ac{TRes-DeltaGRU}-999 estimated on the \texttt{Gem5}-simulated ARM CPU. W12A12 quantization alone reduces energy by 2.8$\times$ versus FP32 while still achieving -54.5\,dBc \ac{ACPR}, and combining it with 56.0\% and 72.5\% sparsity increases the reduction to 4.5$\times$ and 5.2$\times$, respectively, while still meeting a common -45\,dBc \ac{ACPR} target. These results confirm that the active-parameter reduction reported in Table~\ref{tab:optimization} translates into processor-level energy savings. The \texttt{Gem5} statistics also show that memory access dominates the forward-pass energy budget: for the evaluated INT16 cases, L1 D-cache energy is more than an order of magnitude larger than arithmetic energy, and control overhead grows with sparsity. This explains why processor-level savings are smaller than idealized MAC-only estimates, and further motivates joint algorithm-hardware co-design.

\begin{figure}[!t]
    \centering
    \includegraphics[width=1.0\linewidth]{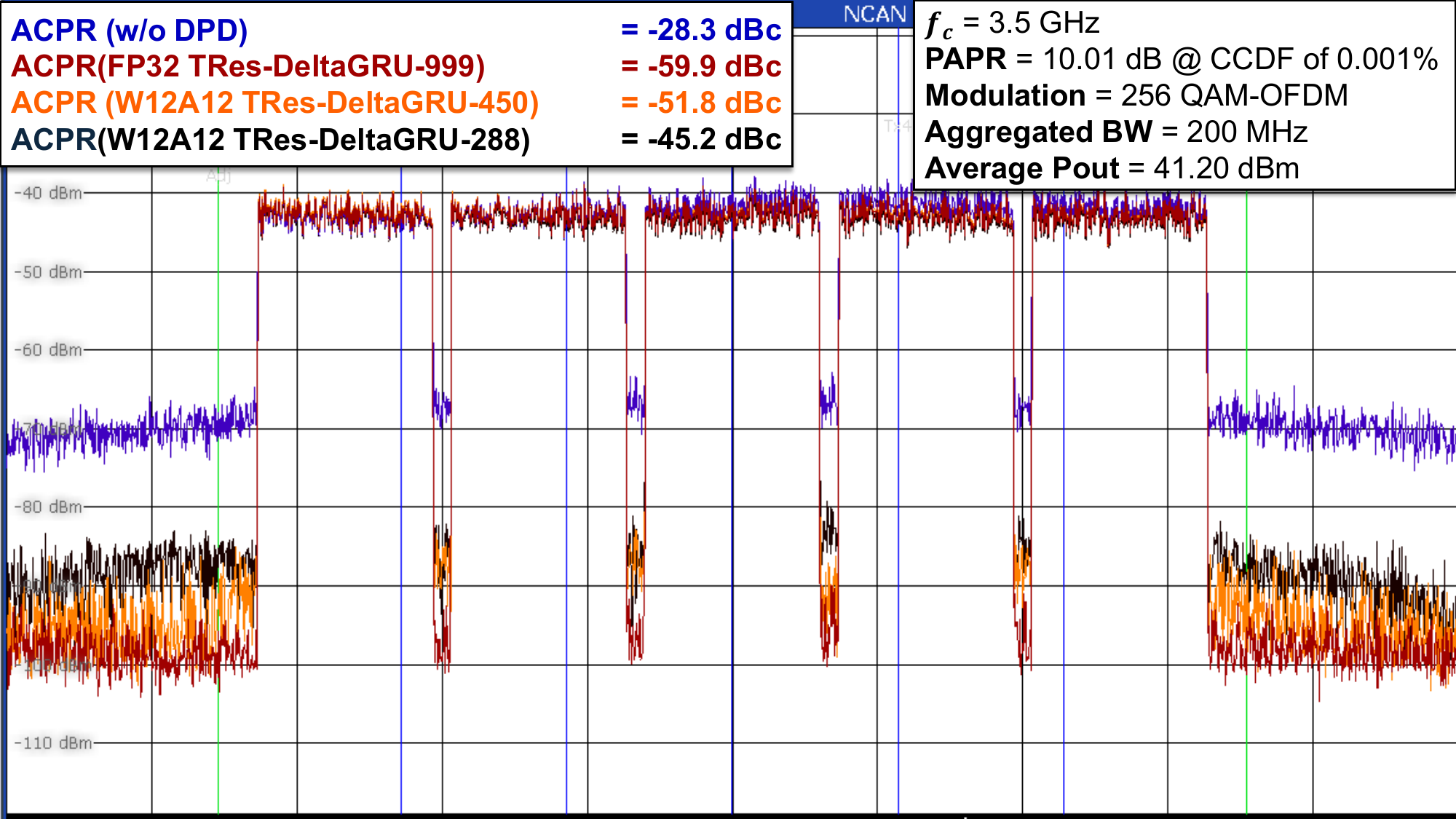}
    \caption{Measured spectrum on the 200 MHz TM3.1a signal without DPD and with \ac{TRes-DeltaGRU} DPD.}
    \label{fig:PSD}
\end{figure}

\begin{figure}[!t]
    \centering
    \includegraphics[width=0.95\linewidth]{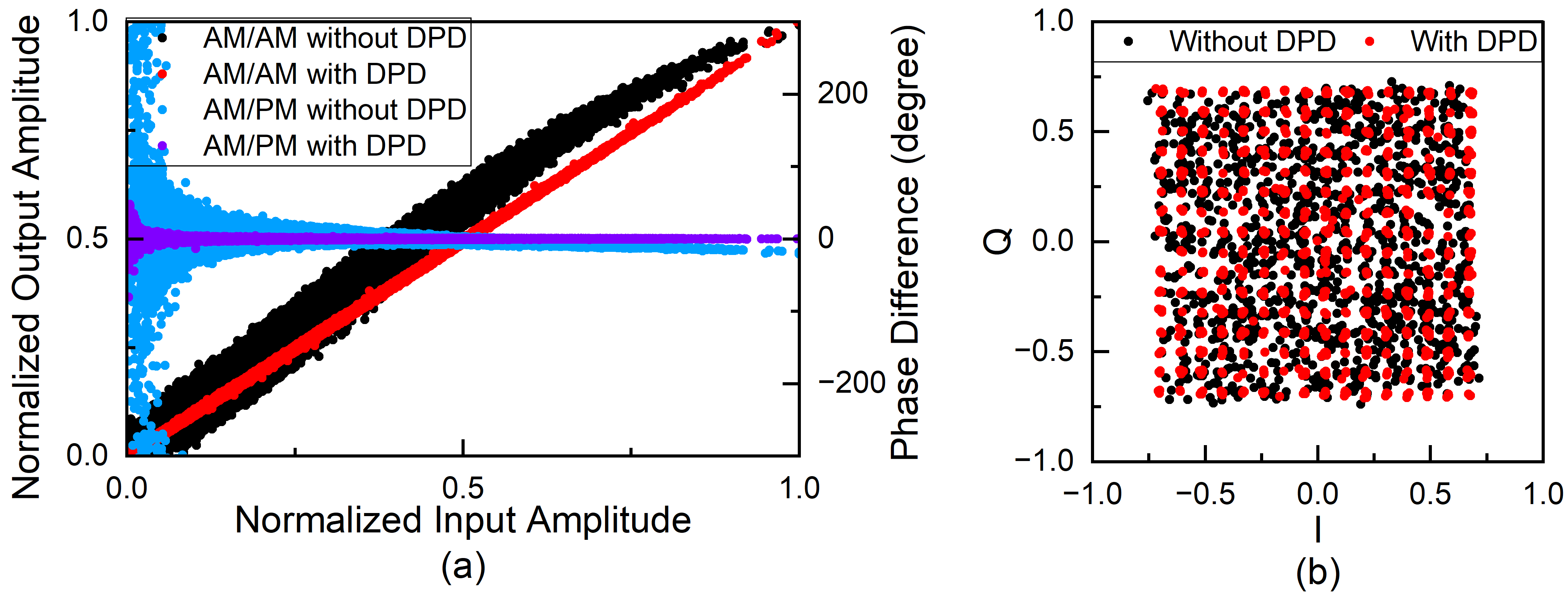}
    \caption{(a) AM/AM and AM/PM characteristic. (b) Constellation on the 200 MHz TM3.1a signal without DPD and with FP32 \ac{TRes-DeltaGRU}-999 DPD.}
    \label{fig:amam}
\end{figure}
Fig.~\ref{fig:PSD} shows representative measured spectra without DPD and with \ac{TRes-DeltaGRU} DPD. Fig.~\ref{fig:amam}(a) shows the corresponding AM/AM and AM/PM characteristics, and Fig.~\ref{fig:amam}(b) shows the constellation without DPD and with the FP32 \ac{TRes-DeltaGRU}-999 model.

\section{Conclusion}
This paper presented \texttt{OpenDPDv2} and \ac{TRes}-\ac{DeltaGRU} for deployment-oriented \ac{NN}-based \ac{DPD}. On a 200\,MHz TM3.1a \ac{OFDM} signal at 41.2\,dBm average output power, FP32 TRes-DeltaGRU-999 achieves -59.9\,dBc \ac{ACPR} and -42.1\,dB \ac{EVM}. At 56.0\% temporal sparsity, the W12A12 model with 450 active parameters still reaches -51.8\,dBc \ac{ACPR} and -35.2\,dB \ac{EVM}, while \texttt{Gem5} indicates 4.5$\times$ forward-pass energy reduction. At 72.5\% sparsity, the estimated reduction reaches 5.2$\times$ while still meeting a common -45\,dBc \ac{ACPR} target. Overall, these results support \ac{TRes}-\ac{DeltaGRU} as a practical building block for efficient wideband \ac{NN}-\ac{DPD}. More importantly, they show that the active-parameter reductions reported in Table~\ref{tab:optimization} are not merely indicators of algorithmic compactness but can translate into meaningful estimated energy savings at the processor level. This makes \texttt{OpenDPDv2} useful not only for model development, but also for deployment-oriented exploration of linearization, sparsity, quantization, and implementation cost within a unified workflow.

\section*{Acknowledgment}
We thank Yi Zhu and John Gajadharsing from Ampleon for lending us the PA DUT and assistance in building the experimental setup. We also thank Prof. Leo C. N. de Vreede from Delft University of Technology and Prof. Anding Zhu from the Chinese University of Hong Kong for related technical discussions and writing suggestions. This work is partially supported by the European Research Executive Agency (REA) under the Marie Skłodowska-Curie Actions (MSCA) Postdoctoral Fellowship program, Grant No. 101107534 (AIRHAR).

\bibliographystyle{IEEEtran}
\bibliography{IEEEabrv,ref}

\end{document}